\documentclass{PoS}

\def\be{\begin{equation}}
\def\ee{\end{equation}}

\def\se1{s$^{-1}$ }

\def\cm2{cm$^2$ }
\def\se1{s$^{-1}$ }

\def\gtsima{$\; \buildrel > \over \sim \;$}
\def\ltsima{$\; \buildrel < \over \sim \;$}
\def\prosima{$\; \buildrel \propto \over \sim \;$}
\def\gsim{\lower.5ex\hbox{\gtsima}}
\def\lsim{\lower.5ex\hbox{\ltsima}}
\def\simgt{\lower.5ex\hbox{\gtsima}}
\def\simlt{\lower.5ex\hbox{\ltsima}}
\def\simpr{\lower.5ex\hbox{\prosima}}

\def\etal{{et al.~}}
\title{Radio emission and jets from Galactic microquasars}

\ShortTitle{Jets from microquasars}

\author{\speaker{Elena Gallo}\\
        Physics Department, University of California Santa Barbara, CA
93106-9530, USA\\
Chandra Fellow\\
        E-mail: \email{elena@physics.ucsb.edu}}

\abstract{Thanks to aggressive campaigns of multi-wavelength observations of
X-ray binaries in outbursts over the last decade or so, we have now reached a
reasonable understanding of their radio phenomenology in response to changes in
the global X-ray properties. Here I shall subjectively review the latest
progresses made in assessing the interplay between inflow and outflow from an
observational point of view, as well as point out a number of open issues that
still need to be addressed both theoretically and observationally. }

\FullConference{VI Microquasar Workshop: Microquasars and Beyond\\
		September 18-22 2006\\
		Societ\`a del Casino, Como, Italy}

\begin{document}
\section{Warning}
As most of our knowledge in this field is based on black hole X-ray binary
(BHB) systems, this review will be inevitably biased towards BHBs'
properties. Section~\ref{nsbh} and part of Section~\ref{oir} will
focus on jets from neutron star X-ray binaries and the importance of comparing
these two classes for a deeper understanding of the jet phenomenon as a
whole.

\section{Radio jets' morphology}

A meaningful description of the different varieties of radio emitting outflows
from BHBs demands a parallel description of the different `X-ray
states'\footnote{Throughout this review, I shall make use of the X-ray state
terminology recently introduced by McClintock \& Remillard~\cite{mccr}.} over
which they are typically observed.  The underlying assumption -- motivated by
the non-thermal spectra, polarization degree and brightness temperature
arguments -- is that of radio emission from X-ray binaries in general as due
to synchrotron radiation from outflowing plasma (although, at very low radio
fluxes, tens of $\mu$Jy, gyrosynchrotron emission from the donor star could
easily dominate).  Radiatively inefficient hard X-ray states are associated
with flat/slightly inverted radio-to-mm spectra and persistent radio flux
levels~\cite{fender01}. In analogy with compact extragalactic radio
sources~\cite{bk79}, the flat spectra are thought to be due to the
superimposition of a number of peaked synchrotron spectra generated along a
conical outflow, or jet, with the emitting plasma becoming progressively
thinner at lower frequencies as it travels away from the jet base.  The jet
interpretation has been confirmed by high resolution radio maps of two hard
state BHBs: Cygnus X-1~\cite{stirling} and GRS~1915+105~\cite{dhawan} are both
resolved into elongated radio sources on milliarcsec scales -- that is tens of
A.U. -- implying collimation angles smaller than a few degrees. Even though no
collimated radio jet has been resolved in any BHB emitting X-rays below a few
per cent of the Eddington limit, it is widely accepted, by analogy with the
two above-mentioned systems, that the flat radio spectra associated with
unresolved radio counterparts of X-ray binaries are originated in conical
outflows. Yet, it remains to be proven whether such outflow would maintain
highly collimated at very low luminosity levels, in the so called `quiescent'
regime. The answer clearly lies in the very jet production mechanism, and its
relation with the inner accretion rate, and will be most likely addressed my
means of magneto-hydrodynamic simulations.

Radiatively efficient, thermal dominant X-ray states, on the contrary, are
associated with no detectable core radio emission~\cite{fender99}; as the
radio fluxes drop by a factor up to 50 with respect to the hard state
(e.g. \cite{cf02},~\cite{corbel04}), this is generally interpreted as the
physical suppression of the jet taking place over this regime.

Transient ejections of optically thin radio plasmons moving away from the
binary core in opposite directions are often observed as a result of bright
radio flares associated with hard-to-thermal X-ray state transitions. This are
surely the most spectacular kind of jets observed from X-ray binaries: those
that have inspired the fortunate term `microquasar'~\cite{mirrod98}. As proven
by the case of GRS~1915+105, and more recently by Cygnus X-1 as
well~\cite{fender06b}, the same source can produce either kind of jets,
persistent/partially self-absorbed, and transient/optically thin, dependently
on the accretion regime.

The above description is meant to give a broad overview of the
different variety of jets powered by BHBs (see~\cite{fender06} for a
thorough review).  Recently, a unified scheme for BHB jets has been
  put forward~\cite{fbg}, whose aim is to provide a more dynamical
  description of a typical BHB cycle, both in terms of accretion and
  jet properties; this will be the subject of the next Section.
\section{A unified scheme for black hole jets}

\begin{figure}
\centering{\includegraphics[width=.9\textwidth]{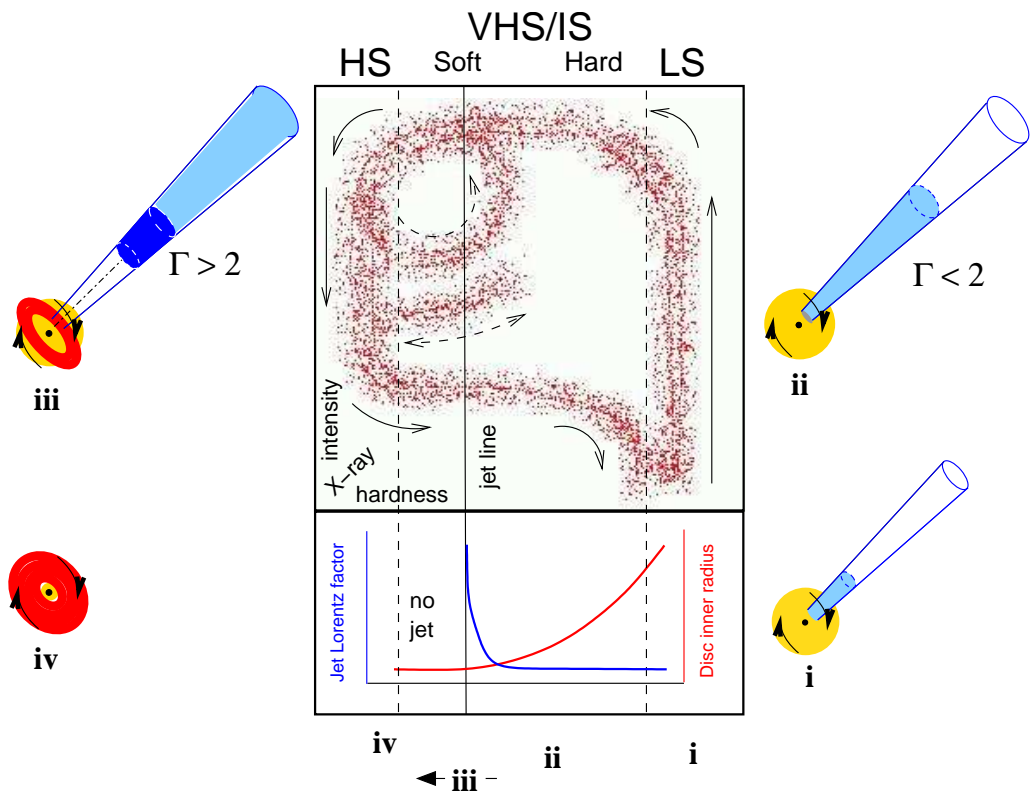}}
\caption{Schematic of the unified model for black hole X-ray binary jets. From
~\cite{fbg}.}
\label{fig:turtle}
\end{figure}
 
Broadly speaking, this phenomenological model aims to put together the various
pieces of a big puzzle that were provided to us by years of multi-wavelength
monitoring of BHBs, and to do so under the guiding notion that the jet
phenomenon has to be looked at as an intrinsic part of the accretion
process. \cite{fbg} collected as many information as possible about the very
moment when major radio flares occur in BHBs, and proposed a way to `read
them' in connection with the X-ray state over which they took place as well as
the observed jet properties prior and after the radio flare itself.  The study
makes use of simultaneous X-ray -- typically RXTE -- and radio -- ATCA and/or
VLA -- observations of four outbursting systems: GRS~1915+105, XTE J1550--564,
GX~339--4 and XTE J1859+229; X-ray Hardness-Intensity Diagrams (HID) have been
constructed for the various outbursts and linked with the evolution of the jet
morphology, radio luminosity, total power, Lorentz factor and so on.

Figure~\ref{fig:turtle}, apparently a.k.a. `the turtle head diagram', 
illustrates a schematic of the model. The top
panel represents a HID for a (rather well
behaved) BHB outburst: the time arrow progresses
counterclockwise. Starting from the bottom right corner, the system is
a low-luminosity hard X-ray state, producing a (supposedly) mildly
relativistic, persistent outflow, with flat radio spectrum.
Its luminosity starts to increase at all wavelengths, while the X-ray
spectrum remains hard; around a few per cent of the Eddington X-ray
luminosity, a sudden transition is made (top horizontal branch) during
which the global properties of the accretion flow change from
radiatively inefficient to efficient (hard-to-thermal dominant state
transition), while a bright radio flare is observed, likely due to a
sudden ejection episode.  This is interpreted as the result of the
inner radius of a geometrically thin accretion disc moving inward, as
illustrated in the bottom panel: the Lorentz factor of the ejected
material, due to the deeper potential well, exceeds that 
of the hard state jet, causing an internal shock to propagate through
it, and to possibly disrupt it. Once the transition to the thermal
dominant state is made, no core radio emission is observed, while
large scale rapidly fading radio plasmons are often seen moving in
opposite direction with highly relativistic speed.

The bright radio flare associated with the transition could coincide with the
very moment in which the hot corona of thermal electrons, responsible for the
X-ray power law in the spectra of hard state BHBs, is accelerated and
ultimately evacuated.  This idea of a sudden evacuation of inner disc material
is not entirely new, and in fact dates back to extensive RXTE/PCA observations
of the rapidly varying GRS~1915+105: despite their complexity, the source
spectral changes could be accounted for by the rapid removal of the inner
region of an optically thick accretion disc, followed by a slower
replenishment, with the time-scale for each event set by the extent of the
missing part of the disc~\cite{belloni97a},\cite{belloni97b}. Subsequently,
multi-wavelength (radio, infrared and X-ray) monitoring of the same source
suggested a connection between the rapid disappearance and follow up
replenishment of the inner disc seen in the X-rays, with the infrared flare
starting during the recovery from the X-ray dip, when an X-ray spike was
observed.

Yet it remains unclear what drives the transition in the radio properties
after the hard X-ray state peak is reached. Specifically, radio observations
of GX~339--4 and XTE J1550-564 and GRS~1915+105 indicate that in this phase
the jet spectral index seems to `oscillate' in an odd fashion, from flat to
inverted to optically thin, as if the jet was experiencing some kind of
instability as the X-ray spectrum softens.  Recent simultaneous RXTE and
INTEGRAL observations of GX~339--4~\cite{belloni06} have shown that the high
energy (few 100s of keV) cutoff typical of hard state X-ray spectra, either
disappears or shifts towards much higher energies { within timescales of hours
($<$8 hr)} during the transition. Previous suggestions of such behaviour were
based on X-ray monitoring campaigns with instruments such as OSSE, for which
the long integration times required in order to accumulate significant
statistics did not allow to constrain the timing and significance of rapid
changes in the X-ray spectra.  The suggestion that the so called `jet line'
(see Figure~\ref{fig:turtle}), where the radio flare is observed and the core
radio emission is suddenly quenched might correspond to a peculiar region in
the time domain -- `The Zone' (see Homan these Proceedings) -- seems to be at
odds with recent radio observations (Fender, again these Proceedings).\\

Finally, there are at least a couple of recent results that might challenge
some of the premises the unified scheme is based on (according to this author
at least). The first one is the notion that, for the internal shock scenario
to be at work and give rise to the bright radio flare at the state transition,
whatever is ejected must have a higher velocity with respect to the
pre-existing hard state steady jet. From an observational point of view, this
was supported, one one side, by the lower limits on the transient jets'
Lorentz factors, typically higher than 2~\cite{fender03}, and, on the other
hand, by the relative small scatter about the radio/X-ray correlation in hard
state BHBs~\cite{gfp} (see Section~\ref{radiox}). The latter has been
challenged on theoretical grounds~\cite{hm04}; while a recent
work~\cite{jmj06} has demonstrated that, from an observational point of view,
the average Lorentz factors do not differ substantially between hard and
transient jets (albeit the estimated Lorentz factors rely on the assumption of
no lateral confinement). While this is further explored in
Section~\ref{speed}, here I wish to stress that, even if the hard and
transient jets' velocities were indeed different, much work needs to be done
in order to test the consistency of the internal shock scenario as a viable
mechanism to account for the observed changes in the radio properties, given
the observational and theoretical constraints for a given source (such as
emissivities, radio/infrared delays, cooling times, mass outflow rates, etc.).

In addition, recent high statistics X-ray observations of a hard state BHB
undergoing outburst~\cite{jm06} suggest that a cool, thin accretion disc
extends already near to the innermost stable circular orbit (ISCO) already
during the bright phases of the hard state, that is prior to the horizontal
brunch in the top panel of Figure~\ref{fig:turtle}.  This would challenge the
hypothesis of a sudden deepening of the inner disc potential well as the cause
of a high Lorentz factor ejection.  Possibly, whether the inner disc radius
moves close to hole prior or during the softening of the X-ray spectrum does
not play such a crucial role in terms of jet properties; if so, then the
attention should be diverted to a different component, such as the
presence/absence, or the size~\cite{homan01}, of a Comptonizing corona (which
could in fact coincide with the very jet base~\cite{mnw05}).\\


One of the most interesting aspects of this proposed scheme -- assuming that
is correct in its general principles -- is obviously its possible application
to super-massive BHs in Active Galactic Nuclei (AGN), and the possibility to
mirror different X-ray binary states into different classes of AGN: radio loud
vs. radio quiet, LLAGN, FRI, FRII etc.. The interested reader is referred to
Jester and Fender, these Proceedings, for novel approaches towards a `great
unification scheme'.

\section{Global correlations}
\subsection{Radio/X-ray}
\label{radiox}
In a first attempt to quantify the relative importance of jet vs. disc
emission in BHBs,~\cite{gfp} collected quasi-simultaneous radio and X-ray
observations of ten hard state sources. This study established the presence of
a tight correlation between the X-ray and the radio luminosity, of the form
$L_{ R}\propto L_{X}^{0.7\pm 0.1}$, first quantified by~\cite{corbel03} for
GX~339--4. The correlation extends over more than 3 orders of magnitude in
$L_{X}$ and breaks down around 2 per cent of the Eddington X-ray luminosity,
above which the sources enter the thermal dominant state, and the core radio
emission drops below detectable levels.  Given the non-linearity of the
correlation, the ratio radio-to-X-ray luminosity increases towards
quiescence; one wonders however whether the steady jet survives in this very
low luminosity state (with $L_{\rm X}\simlt 10^{33.5}$ erg sec$^{-1}$,
i.e. below a few $10^{-5} L_{\rm Edd}$). In such a regime, very few systems
have been detected in the radio band, mainly because of sensitivity
limitations on the existing telescopes.  Given the quite large degree of
uncertainty about the overall structure of the accretion flow in quiescence,
it has even been speculated that the total power output of quiescent BHBs
could be dominated by a radiatively inefficient outflow~\cite{fgj}, rather
than by the local dissipation of gravitational energy in the accretion flow.

Due to its extremely low X-ray luminosity ($L/L_{Edd}\sim10^{-8.5}$) and
relative proximity, the 10 solar mass BH in A0620--00 represents the most
suitable known system to probe the radio/X-ray correlation beyond the hard
state.  Deep VLA observations of this system, performed in 2005 August,
resulted in the first radio detection of a quiescent BHB emitting at such low
X-ray luminosities.  The level of radio emission -- 51 $\mu$Jy at 8.5 GHz --
is the lowest ever measured in an X-ray binary. At a distance of 1.2 kpc, this
corresponds to a radio luminosity $L_{\rm R}=7.5\times 10^{26}$ erg
sec$^{-1}$. By analogy with higher luminosity systems, partially self-absorbed
synchrotron emission from a relativistic outflow appears to be the most likely
interpretation. Free-free wind emission is ruled out on the basis that far too
high mass loss rates would be required, either from the companion star or the
accretion disc, to produce observable emission at radio wavelengths, while
gyrosynchrotron radiation from the corona of the companion star is likely to
contribute to less than 5 per cent to the measured flux density.

The simultaneous \emph{Chandra} observation allowed to test and extend the
radio/X-ray correlation for BHBs by 3 orders of magnitude in $L_{\rm X}$.  The
measured radio/X-ray fluxes confirm the validity of a non-linear scaling
between the radio and X-ray luminosity in hard and quiescent systems; with the
addition of the A0620--00 point, {$L_{\rm R}\propto L_{\rm X}^{0.58\pm0.16}$
provides a good fit to the data for $L_{\rm X}$ spanning between $10^{-8.5}$
and $10^{-2} L_{\rm Edd}$ (see Figure~\ref{fig:gallo06}). The fitted slope,
albeit consistent with the previously reported value of $0.7\pm0.1$, is
admittedly affected by the uncertainties in the distance to GX~339--4, for
which the correlation extends over 3 orders of magnitude in $L_{\rm X}$ and
holds over different epochs. Pending a more accurate determination of the
distance to this source, we can nevertheless exclude the relation breaking
down and/or steepening in quiescence. That the non linear radio/X-ray
correlation for hard state black hole X-ray binaries extends down to very low
quiescent luminosities implies that the ratio of jet-to-accretion radiative
power is a decreasing function of $L_X$ over the luminosity range that is
explorable with current instrumentation~\cite{gallo06}.

\subsection{Optical-IR/X-ray}
\label{oir}
\begin{figure}
\centering{\includegraphics[width=.775\textwidth]{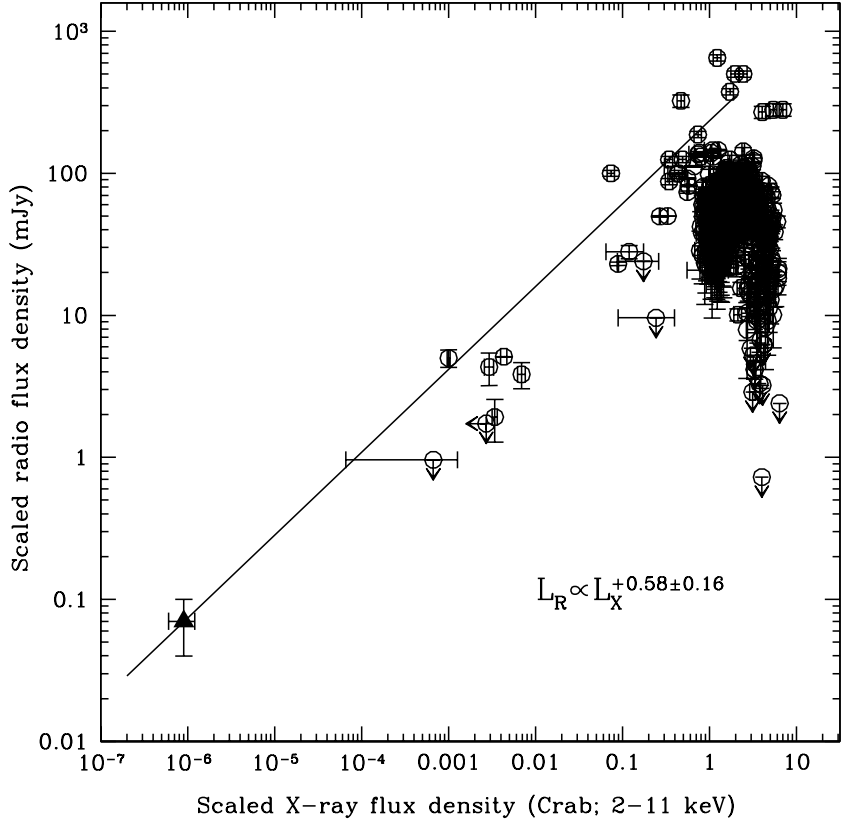}}
\caption{Radio/X-ray correlation for black hole X-ray binaries, with the
addition of the $L_X/L_{Edd}\simeq 10^{-8.5}$ BH A0620--00. From~\cite{gallo06}.}
\label{fig:gallo06}
\end{figure}

In order to quantify the relative importance of jet vs. disc emission as a
function of the state and luminosity in a frequency window that is undoubtedly
crucial and yet shaped by a number of competing mechanisms,
\cite{russell} put together nearly-simultaneous optical-IR (termed
OIR) and X-ray observations of X-ray binaries, BHs and neutron stars.  A
global correlation is found between OIR and X-ray luminosity for low-mass BHBs
in the hard state, of the form $L_{OIR}\propto L_X^{0.6}$ (see
Figure~\ref{fig:opt}).  This correlation holds over eight orders of magnitude
in $L_X$ and includes data from BHBs in quiescence. A similar correlation is
found in low-mass neutron star X-ray binaries in the hard state. Similarly to
what happens for the radio emission, for thermal dominant state BHBs all of
the near-IR and some optical emissions are suppressed, indicating that the jet
is quenched during the hard-to-thermal transition. By comparing these
empirical correlations with existing models, \cite{russell} come to the
conclusion that, for the BHs, X-ray reprocessing in the disc and emission from
the jets both contribute to the optical-IR while in the hard state, with the
jet accounting for up to 90 per cent of the near-IR emission.  In addition, it
is shown that the optically thick jet spectrum of BHBs is likely to extend to
near the K band.

X-ray reprocessing dominates the in hard state neutron stars, with
possible contributions from the jets and the
viscously heated disc, only at high luminosities.
\section{Black holes vs. neutron stars}
\label{nsbh}
The mechanism(s) of jet production, from an {\it observational} point of view,
remains essentially unconstrained. While in the case of super-massive BHs in
AGN it is often implicitly assumed that the jets extract their energy from the
rotation of the centrally spinning black hole via large scale magnetic field
lines that thread the horizon, in the case of X-ray binaries, the relatively
low (lower limit on the) jets' Lorentz factor do not appear to require
especially efficient launching mechanisms. On the `experimental' side,
substantial improvements are being made with fully relativistic
magneto-hydrodynamic simulations (see Krolik, these Proceedings); from the
observer perspective it seems (to this author) that a fruitful -- and yet
relatively unexplored -- path to pursue is that to compare in a systematic
fashion the properties of jets in black hole systems to that of e.g. low
magnetic field neutron stars. A comprehensive study comparing the radio
properties of BHs and neutron stars~\cite{mf06} has highlighted a number of
relevant difference/similarities (see Figure~\ref{fig:ns}):
\begin{enumerate}
\item
Below a few per cent of the Eddington luminosity (in the hard, radiatively
inefficient states) both black holes and neutron stars produce steady compact jets,
while transient jets are associated with variable sources/flaring activity at
the highest luminosities.  
\item
For a given X-ray luminosity, the neutron stars are less radio loud, typically
by a factor of 30.
\item
Unlike black holes, neutron stars do not show a strong
suppression of radio emission in the soft states.
\item
Hard state neutron stars exhibit a much steeper correlation between radio and
X-ray luminosities (see K\"ording, these Proceedings for a
theoretical interpretation),
\end{enumerate}

One other difference, even though it should be confirmed by observations of a
larger sample of neutron stars, has to due with the location of the optically
thick-to-thin jet break.  While the study by Russell \etal~\cite{russell}
indicate that, for the BHBs, the break takes place in the mid-IR, this could
happen at lower frequencies for the neutron stars (we know however from
observations of GX~339--4, the only BHB where the optically thin jet spectrum
has been perhaps observed~\cite{cf02}, that the exact frequency of the break
can vary with the overall luminosity, possibly reflecting changes in the
magnetic field energy density, particle density and mass loading at the jet
base).
That the optically thin jet IR-emission in GX~339--4 connects smoothly with
the hard X-ray power law has led to challenge the `standard' Comptonization
scenario for the hard X-ray state~\cite{mff01}, whereas recent Spitzer
observations of the ultra-compact neutron star X-ray binary 4U~0614+091, with
the Infrared Array Camera, unambiguously showed that jet the break frequency
must take place in the far-IR in this system, effectively ruling out a
synchrotron origin for the X-ray power law~\cite{migliari06}.  The upper limit
on the break frequency immediately allows to conclude that, at least in terms
of radiative output, the jet power in this neutron star X-ray binary is lower
than in a hard state BHB emitting X-rays at a comparable level, at least by a
factor of ten.

Nevertheless, perhaps ironically, the most relativistic jet discovered in the
Galaxy so far, is that the neutron star X-ray binary Circinus
X-1~\cite{fendercirx1}, for which the inferred Lorentz factor exceeds 15.
\begin{figure}
\hspace{1cm}
\includegraphics[width=.95\textwidth]{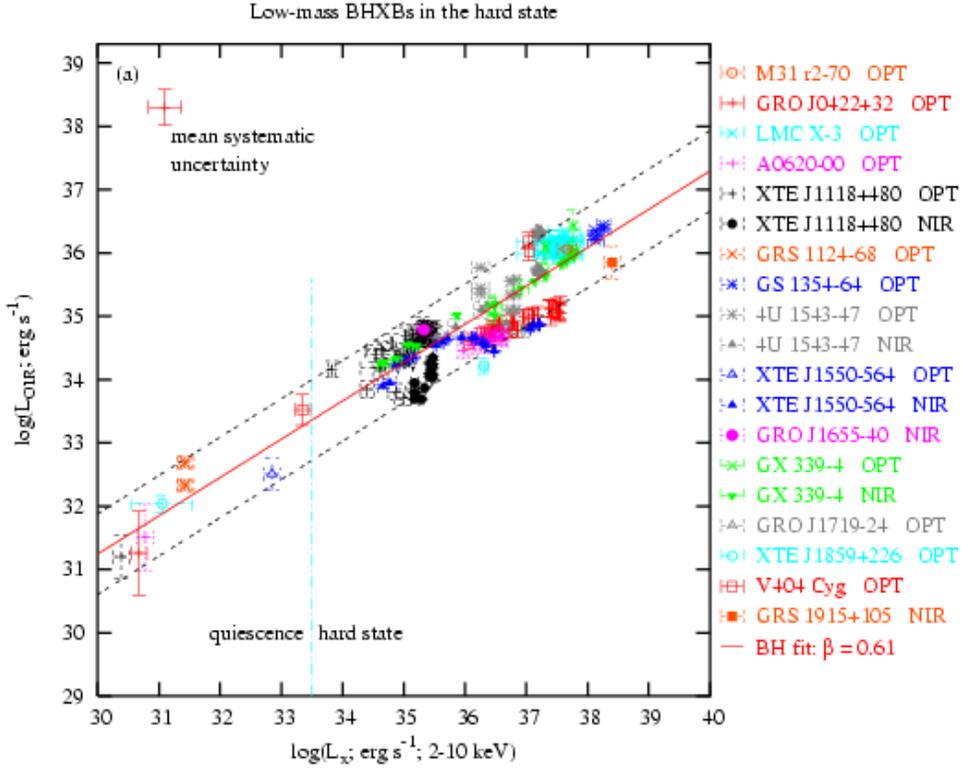}
\caption{Optical-IR vs. X-ray correlation in hard state black holes. From~\cite{russell}}.
\label{fig:opt}
\end{figure}

\begin{figure}
\centering{\includegraphics[width=.85\textwidth]{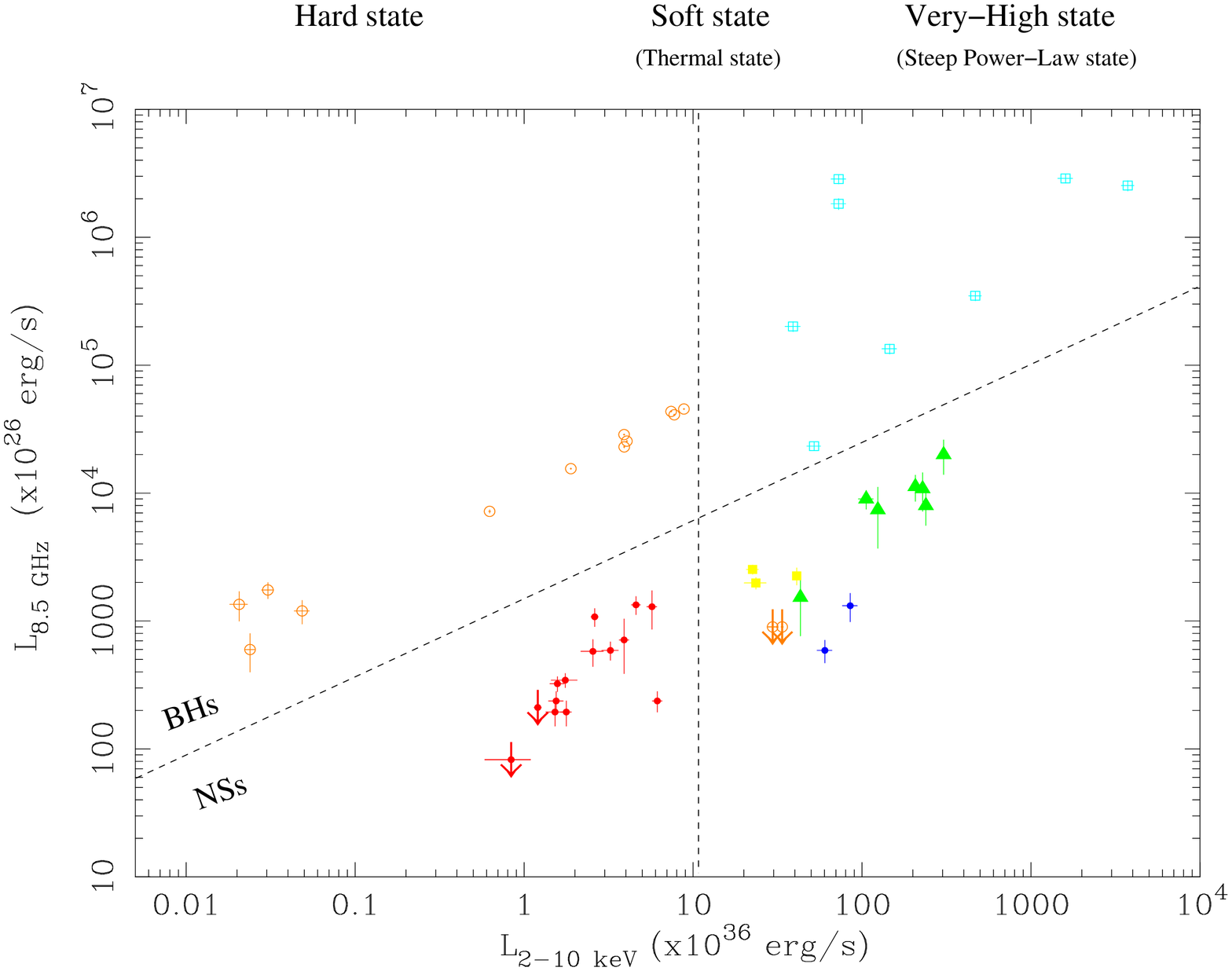}}
\caption{A comparison between radio/X-ray properties of neutron star and black
hole X-ray binaries. From~\cite{mf06}.}
\label{fig:ns}
\end{figure}
\section{Jet power}
Estimate of the radiative jet power in X-ray binaries are severely biased by
the contamination of competing emission mechanisms in the IR-optical band,
most notably the donor star and the outer accretion disc, where the break from
partially self-absorbed to optically thin is thought to occur.  Given that
most of the jet radiation is emitted at higher frequencies, the jet `radiative
efficiency' depends ultimately on the location of the high-energy cutoff
induced by the higher synchrotron cooling rate of the most energetic
particles. Once again, this quantity has proved hard to measure.

A fruitful method, again borrowed from the AGN community, is that to constrain
the jet power-times-lifetime product by looking at its interaction with the
surrounding interstellar medium (see Heinz, these Proceeding for a
comprehensive review). A well known case is that of the nebula around the
first 
Galactic jet source discovered: SS~433. The `ears' of W50
act as an effective calorimeter for the jets' mechanical power, which is
estimated to be greater than $10^{39}$ erg sec$^{-1}$ (e.g.~\cite{begelman80}).
More recently, a low surface brightness arc of radio emission has been 
discovered around Cygnus X-1~\cite{gallo05} (Figure~\ref{fig:arc}) and interpreted in terms of a shocked
compressed hollow sphere of free-free emitting gas driven by an under-luminous
synchrotron lobe inflated by the jet of Cygnus X-1.
\begin{figure}
\hspace{-0.5cm}
\centering{\includegraphics[width=.785\textwidth,angle=270]{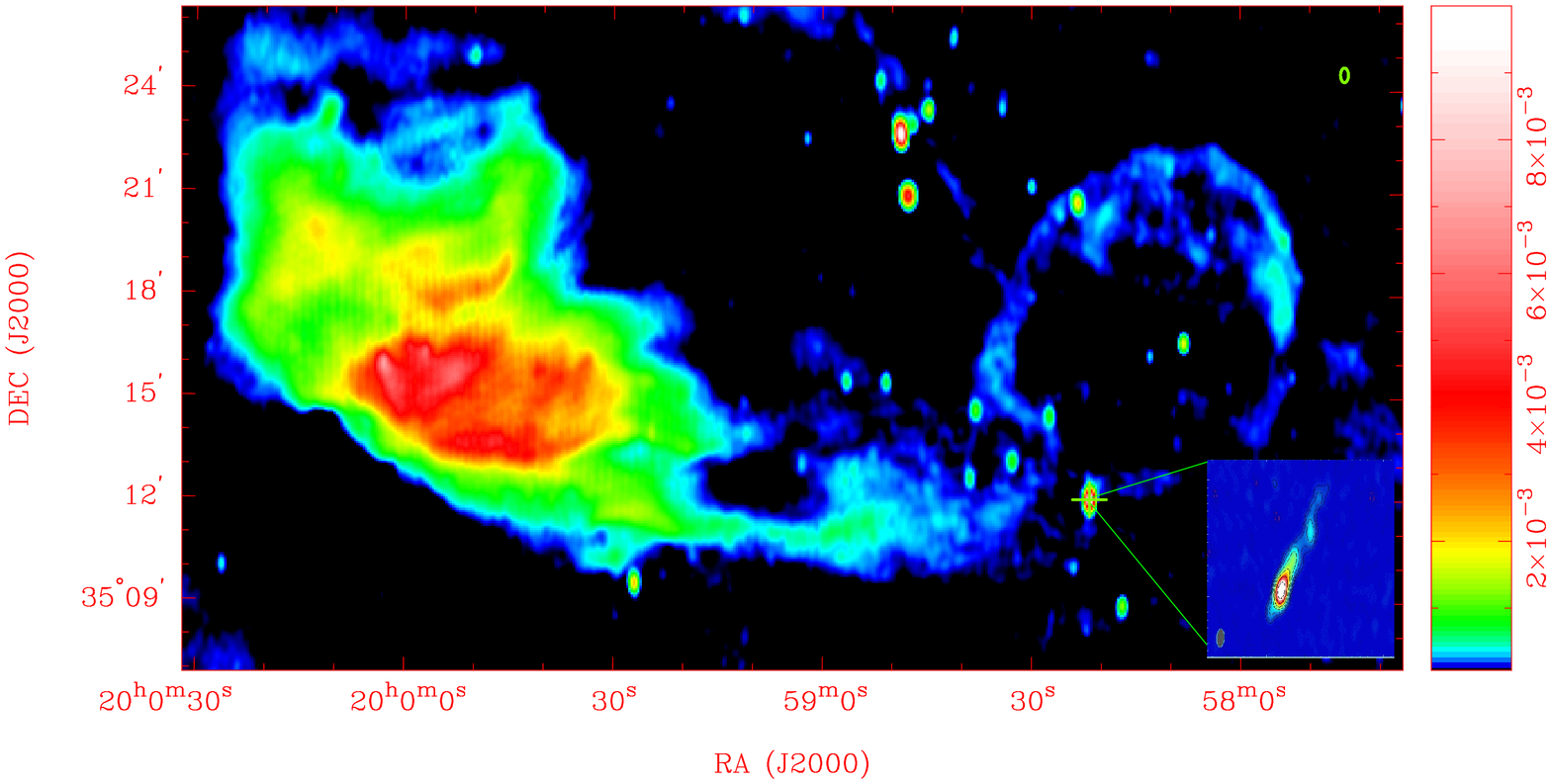}}
\caption{A jet-powered nebula around Cygnus X-1. From~\cite{gallo05}.}
\label{fig:arc}
\end{figure}
The lack of a visible counter arc is ascribed to the lower
interstellar matter density in the opposite direction; in fact, it is
likely that X-ray binary jets require a particularly dense environment
in order to produce visible signs of interaction with the
surroundings (Heinz, these Proceedings). The reader is referred to
Russell, these Proceedings, for follow-up optical observations of the
Cygnus X-1 nebula, as well as other X-ray binaries', and Tudose, these
Proceedings, for a study of the jet-powered radio nebula around the
neutron star X-ray binary Circinus X-1~\cite{tudose06}.\\

Estimates of the total jet power are obviously to be compared with the total
accretion energy budget; this is especially interesting in very low-luminosity
systems, where the fate of the accreting gas remains a matter of debate.
Observations of highly sub-Eddington BHs, most notably the Galactic Centre
super-massive BH, paved the way to radiatively inefficient accretion flow
models.  By reviewing the vast literature on the subject, one immediately
comes to the conclusion that the most widely accepted/adopted model for
reproducing the spectral energy distribution of quiescent BHs is the
`advection-dominated accretion flow' solution (ADAF; e.g.~\cite{ny94}). Here,
a significant fraction of the viscously dissipated energy remains locked up in
the gas as heat, and is advected inward.  The ADAF model successfully accounts
for the overall shape of the UV-optical-X-ray spectra of quiescent BHBs (see
e.g.~\cite{mcc03} for an application to the high quality data of XTE
J1118+480).  Nevertheless, alternative suggestions are well worth being
considered.
\cite{bb99} elaborated an `adiabatic inflow-outflow solution', in
which the excess energy and angular momentum is lost to an outflow at all
radii; the final accretion rate into the hole may be only a tiny fraction of
the mass supply at large radii.

A0620--00 provides an instructive (albeit not conclusive) test for those
models. Interestingly enough, based on models for the optical/UV emission of
the outer accretion disc in dwarf novae, corrected downward to account for the
mass difference,
\cite{mcc95} estimate $\dot M_{\rm out} = y 10^{-10} M_{\odot}$ yr$^{-1}$ for
A0620--00, where $y$ is a factor of the order unity, that can be
up to a few. The putative luminosity associated with $\dot M_{\rm out}$, {if
it was to reach the hole with a standard radiative efficiency of 10 per
cent}, would be $L_{ tot}
\equiv \eta \dot M_{out}c^2 \simeq 6 \times 10^{35} y~(\eta / 0.1)$ erg
sec$^{-1}$, about 5 orders of magnitude larger than the measured X-ray
luminosity (where $\eta$ is the accretion efficiency, which depends only on
the BH spin).  The various radiatively inefficient accretion flow models offer
different explanations for the much lower luminosities that are observed in
terms of different `sinks' for the energy.  Making use the expression for the
jet radiative efficiency by Heinz \& Grimm~\cite{heinzgrimm}, it can be shown
directly that, in the case of A0620--00, the outflow kinetic power accounts
for a sizable fraction of the accretion energy budget, and thus must be
important with respect to the overall accretion dynamics of the system. In
spite of the many uncertainties in the above calculations, this would
effectively rule out a {\em pure} ADAF solution for the dynamics of the
accretion flow in quiescence. However, within these uncertainties there is
still room for a hybrid solution to apply, one in which at each $\dot{M}$
about half of the energy is carried away by the outflow, while the rest is
advected inward and finally added to the BH mass. It is worth mentioning that
the possibility that an ADAF could naturally launch outflows was already
explored back in 1995~\cite{ny95}, even though the outflow contribution to the
radiation spectrum is typically neglected in the calculations.

\section{Jet composition, speed and confinement}
\label{speed}
A less debated issue seems that of X-ray binary jets' speed. Perhaps
erroneously so. It is often
assumed that the velocity the the steady jet is only mildly
relativistic, with $\Gamma \simeq 2$~\cite{gfp}. This comes from the
relative spread about the radio/X-ray correlation, interpreted as
evidence for a low average Lorentz factor. However, as mentioned in
Section~\ref{radiox}, this argument has been confuted on theoretical
grounds~\cite{hm04}.  As far as the transient jets are concerned,
there is a high degree of uncertainties in estimating their Lorentz
factors, mainly because of distance uncertainties~\cite{fender03}.

In a recent work~\cite{jmj06} make a substantial step forward by means of the
observational upper limits on the jets' opening angles.  This method
relies on the fact that, while the jets could undergo transverse expansion at
a significant fraction of the speed of light, time dilation effects associated
with the bulk motion will reduce their apparent opening angles.  \cite{jmj06}
have calculated the Lorentz factors required to reproduce the small opening
angles that are observed in most X-ray binaries, with very few exceptions,
{under the crucial assumption of no confinement}.  The derived values, mostly
lower limits, are larger than typically assumed, with a mean $\Gamma$>10 (see
Figure~\ref{fig:jmj}). No systematic difference appears to emerge between hard
state steady jets and transient plasmons.  If indeed the transient jets were
as relativistic as the steady jets, as already mentioned, this would challenge
the hypothesis of internal shocks at work during hard-to-thermal state
transitions in BHBs. In order for that scenario to be viable, the transient
jets must have higher Lorentz factor; in other words, steady jets ought to be
laterally confined.\\

The issue of the jets' matter content remains highly debated. Perhaps with the
exception of SS~433 (e.g. \cite{watson},~\cite{mfm02}), where atomic lines
have been detected at optical at X-ray wavelengths along the jets, various
studies come to different conclusions for different sources. For example,
circular polarization, which in principle can provide an excellent tool for
investigating the baryonic content of the jets, is only detected in a handful
of sources (see~\cite{fender06} and references therein), where no strong
conclusion could be placed yet.  Entirely different studies (e.g. based on
modelling large scale jet-ISM interaction structures by means of self-similar
jet fluid models) also draw different conclusions: in the case of Cygnus X-1
for instance, some authors (\cite{gallo05} and ~\cite{heinz06}) argue for cold
baryons in the flow, while an electron/positron jet seems to be favored in GRS
1915+105 based on energetics arguments~\cite{fenderpol}.

\begin{figure}
\centering{\includegraphics[width=.65\textwidth]{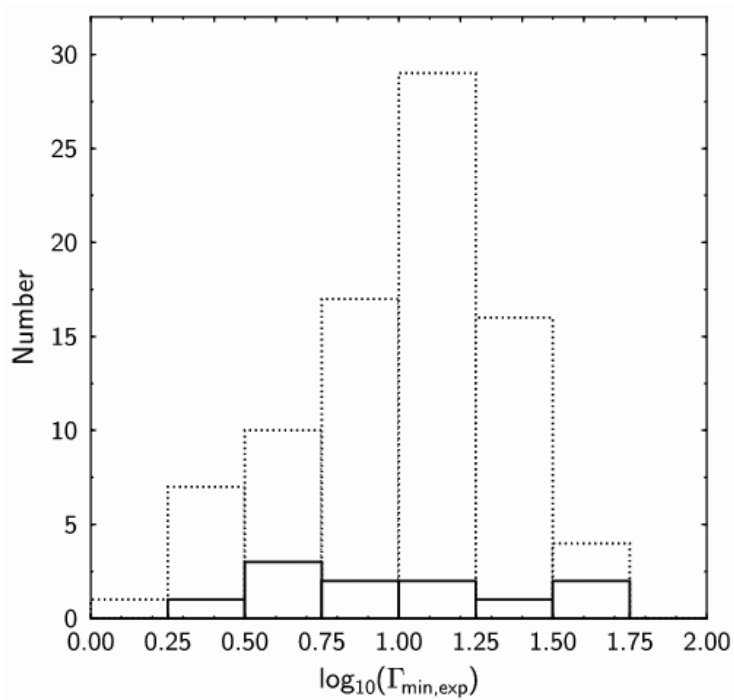}}
\caption{Mean Lorentz factors for X-ray binaries from opening angles'
constraints (solid), compared to AGN from proper motions'
(dashed). From~\cite{jmj06}.}
\label{fig:jmj}
\end{figure}
\section{Open issues}
I shall conclude this review by simply listing a number of issues that remain
open in this field, and that may stimulate further thoughts.
\begin{itemize}
\item Unified model refinement: while it seems to hold in its basic
principles, the unified scheme for BHB jets needs to be carefully tested on
theoretical grounds and hopefully answer the questions posed by recent
observations, such as the possibility that a thin disc is already present down
to the ISCO prior to the actual hard-to-thermal state transition. If so, what 
causes the bright radio flare? 
In addition: under the internal shock scenario working-hypothesis, would the
observed radio plasmons be able to radiate synchrotron emission at thousands of
A.U. from the actual location of the shock? If so, what are the required
initial conditions at the jet base? If not, are perhaps multiple or external
shocks required?
\item Neutron star jets: how do they behave in detail? Are they truly less
powerful 
with respect to BH jets? Are they collimated (no elongated radio structure has
been resolved yet in a `hard state' neutron star)?  In order to complete the
picture and possibly gain meaningful information on the very jet formation
mechanism(s), aggressive multi-wavelength campaigns are needed for neutron
stars as well (and are already in place).
\item The big picture: the ultimate goal is of course that to bridge
the gap between micro-and-macro: between stellar and super-massive BHs. It
seems that we are now at the beginning of a new phase in our understanding of
the behaviour of matter in the vicinity of black holes, one in which
quantitative scalings have emerged on all mass scales, and we may be able to
apply the results obtained from BHBs to the study of AGN, and vice versa (see
Jester and Fender, these Proceedings). Eventually, our relatively deep
knowledge about BHB radio/X-ray cycles could help us understanding AGN duty
cycles and radio modes, which in turn appear to play a role of paramount
importance in 
driving the very evolution of galaxies over cosmological times.
 
\end{itemize}

\section*{Acknowledgments}
I wish to thank a number of collaborators whose research has largely
contributed to shape this review, and in particular Rob Fender,
James Miller-Jones and Dave Russell. Support to this work is provided through 
Chandra Postdoctoral Fellowship Award PF5-60037, issued by the Chandra
X-Ray Observatory Centre under NASA contract NAS8-03060.


\begin{thebibliography}{99}

\bibitem{belloni06}
Belloni T. M. \etal, \emph{INTEGRAL/RXTE high-energy observation of a state transition of GX 339-4} (2006) \emph{MNRAS} {\bf 367} 1113

\bibitem{belloni97a}
Belloni T. M., M\'{e}ndez M., King A. R., van der Klis M., van Paradijs J.,
\emph{A Unified Model for the Spectral Variability in GRS 1915+105} (1997)
\emph{ApJ} {\bf 488} {L109} 

\bibitem{belloni97b}
Belloni T. M., M\'{e}ndez M., King A. R., van der Klis M., van Paradijs J.,
\emph{An Unstable Central Disk in the Superluminal Black Hole X-Ray Binary GRS
1915+105} (1997) {\bf 479} L145

\bibitem{begelman80} Begelman M. C., King A. R., Pringle J. E.,
  \emph{The nature of SS~433 and the ultraluminous X-ray sources}
  (1980) \emph{MNRAS} {\bf 370} 399

\bibitem{bb99}
Blandford R. D., Begelman M. C., \emph{On the fate of gas accreting at a low rate on to a black hole} (1999) \emph{MNRAS} {\bf 303} L1

\bibitem{bk79}
Blandford R. D., K\"onigl A., \emph{Relativistic jets as compact radio sources}  (1979) \emph{ApJ} {\bf 232} 34

\bibitem{cf02} Corbel S., Fender R. P., \emph{Near-Infrared Synchrotron
Emission from the 
Compact Jet of GX 339-4} (2002) \emph{ApJ} {\bf 573} L35

\bibitem{corbel04} Corbel S., Fender R. P., Tomsick J. A., Tzioumis A. K.,
Tingay S., \emph{On the Origin of Radio Emission in the X-Ray States of XTE
J1650-500 during the 2001-2002 Outburst} (2004) \emph{ApJ} {\bf 617} 1272

\bibitem{corbel03} Corbel S., Nowak M., Fender R. P., Tzioumis A. K.,
  Markoff S., \emph{Radio/X-ray correlation in the low/hard state of
    GX 339-4} (2003) {A\&A} {\bf 400} 1007

\bibitem{dhawan} Dhawan V., Mirabel I. F., Rodr\'\i guez L. F.,
  \emph{AU-Scale Synchrotron Jets and Superluminal Ejecta in GRS
    1915+105} (2000) \emph{ApJ} {\bf 543} 373

\bibitem{fender06} Fender R. P., \emph{Compact Stellar X-Ray Sources}, in 
  Lewin W. H. G., van der Klis M. eds, Cambridge Univ. Press,
  Cambridge (2006)

\bibitem{fender03} Fender R. P., \emph{Uses and limitations of relativistic jet proper motions: lessons from Galactic microquasars
} (2003) \emph{MNRAS} {\bf 340} 1353

\bibitem{fender01} Fender R. P., \emph{Powerful jets from black hole
  X-ray binaries in low/hard X-ray states} (2001) \emph{MNRAS} {\bf
  322} 31

\bibitem{fender06b} Fender R. P. \etal \emph{A transient relativistic
  radio jet from Cygnus X-1} (2006) \emph{MNRAS} {\bf 369} 603


\bibitem{fbg} Fender R. P., Belloni T., Gallo E., \emph{Towards a
  unified model for black hole X-ray binary jets} (2004) \emph{MNRAS}
  355, 1105

\bibitem{fendercirx1} Fender R. P., Wu K., Johnston H., Tzioumis T.,
  Jonker P. G., Spencer R., van der Klis M., \emph{An
    ultra-relativistic outflow from a neutron star accreting gas from
    a companion} (2004) \emph{Nature} {\bf 427} 222


\bibitem{fgj}
Fender R. P., Gallo E., Jonker P. G., \emph{Jet-dominated states: an alternative to advection across black hole event horizons in `quiescent' X-ray binaries} (2003) \emph{MNRAS} {\bf 343} L99
	
\bibitem{fenderpol}
Fender R. P., Rayner D., Trushkin S. A., O'Brien K., Sault R. J., Pooley
G. G., Norris R. P., \emph{Variable circular polarization associated with relativistic ejections from GRS 1915 + 105} (2002) \emph{MNRAS} {\bf 330} 212

\bibitem{fender99} Fender R. P. \etal, \emph{Quenching of the Radio
  Jet during the X-Ray High State of GX 339-4} (1999) \emph{ApJ} {\bf
  519} L165

\bibitem{gallo06} Gallo E. \etal, \emph{A radio-emitting outflow in the quiescent state of A0620-00: implications for modelling low-luminosity black hole binaries}  (2006) \emph{MNRAS} {\bf 370}
  1351 

\bibitem{gallo05} Gallo E. \etal, \emph{A dark jet dominates the power
  output of the stellar black hole Cygnus X-1} (2005) \emph{Nature}
  {\bf 436} {819}

\bibitem{gfh}
Gallo E., Fender R. P., Hynes R. I., \emph{The radio spectrum of a
  quiescent stellar mass black hole} 2005, MNRAS, 356, 1017 


\bibitem{gfp} Gallo E., Fender R. P., Pooley G. G., \emph{A universal
  radio/X-ray correlation in hard state black hole binaries} (2003)
  \emph{MNRAS} {\bf 344} 60


\bibitem{heinz06}
Heinz S., \emph{Composition, Collimation, Contamination: The Jet of Cygnus X-1} (2006) \emph{ApJ} {\bf 636} 316

\bibitem{heinzgrimm} Heinz S., Grimm H. J., \emph{Estimating the Kinetic Luminosity Function of Jets from Galactic X-Ray
Binaries} (2005) \emph{633} 384

\bibitem{hm04}
Heinz S., Merloni A., \emph{Constraints on relativistic beaming from estimators of the unbeamed flux} (2004) \emph{MNRAS} {\bf 355} L1 

\bibitem{homan01}
Homan J., Wijnands R., van der Klis M., Belloni T., van Paradijs J.,
Klein-Wolt M., Fender R., M\'{e}ndez M., \emph{Correlated X-Ray Spectral and
Timing Behavior of the Black Hole Candidate XTE J1550-564: A New
Interpretation of Black Hole States} (2001) \emph{ApJ} {\bf 132} 377 

	
\bibitem{mnw05}
Markoff S., Nowak M. A., Wilms J., \emph{Going with the Flow: Can the Base of
Jets Subsume the Role of Compact Accretion Disk Coronae?} (2005) \emph{ApJ}
{\bf 635} 1203


\bibitem{mff01}
Markoff S., Falcke H., Fender R., \emph{A jet model for the broadband spectrum of XTE J1118+480. Synchrotron emission from radio to X-rays in the Low/Hard spectral state} (2001) \emph{A\&A} {\bf 372} L25

\bibitem{mccr} McClintock J. E., Remillard R. A., \emph{Compact
  Stellar X-Ray Sources}, in Lewin W. H. G., van der Klis M., eds,
  Cambridge Univ. Press, Cambridge (2006)

\bibitem{mcc03} McClintock J. E. \etal, \emph{Multi-wavelength Spectrum of the
Black Hole XTE J1118+480 in Quiescence} (2003) \emph{ApJ} {\bf 593} {435}

\bibitem{mcc95} McClintock J. E., Horne K., Remillard R. A., \emph{The DIM
inner accretion disk of the quiescent black hole A0620-00} (1995) \emph{ApJ}
{\bf 442} 358



\bibitem{jm06} Miller J. M. \etal, \emph{A Long, Hard Look at the
  Low-Hard State in Accreting Black Holes}, submitted to \emph{ApJ},
  [astro-ph/0602633]

\bibitem{jmj06}
Miller-Jones J. C. A., Fender R. P., Nakar E., \emph{Opening angles, Lorentz factors and confinement of X-ray binary jets} (2006) \emph{MNRAS} {\bf 367} 1432

\bibitem{migliari06}
Migliari S. \etal, \emph{Spitzer Reveals Infrared Optically Thin Synchrotron
Emission from the Compact Jet of the Neutron Star X-Ray Binary 4U 0614+091}
(2006) \emph{ApJ} {\bf 643} L41 

\bibitem{mf06} Migliari S. \& Fender R. P., \emph{Jets in neutron star
  X-ray binaries: a comparison with black holes} (2006) \emph{MNRAS}
  {\bf 366} 79

\bibitem{mfm02} Migliari S., Fender R. P., M\'{e}ndez, \emph{Iron
  Emission Lines from Extended X-ray Jets in SS~433: Reheating of
  Atomic Nuclei} (2002) \emph{Science} {\bf 297} 1673

\bibitem{mirrod98b}
Mirabel I. F., Rodr\'\i guez L. F., \emph{Accretion instabilities and jet formation in GRS 1915+105} (1999) \emph{ARA\&A} {\bf 37} 409

\bibitem{mirrod98}
Mirabel I. F., Rodr\'\i guez L. F., \emph{Microquasars in our Galaxy} (1998) \emph{Nature} {\bf 392} 673  

	

\bibitem{ny94} Narayan R., Yi I., \emph{Advection-dominated accretion: A
self-similar solution} (1994) \emph{ApJ} {\bf 428} L13

\bibitem{ny95} Narayan R., Yi I., \emph{Advection-dominated accretion: Self-similarity and bipolar outflows} (1995) \emph{ApJ} {\bf 444} 231

\bibitem{russell} Russell D. M. \etal, \emph{Global
  optical/infrared-X-ray correlations in X-ray binaries: quantifying
  disc and jet contributions}, (2006) \emph{MNRAS} {\bf 371} {1334}.


\bibitem{stirling} Stirling A. M. \etal, \emph{A relativistic jet from
  Cygnus X-1 in the low/hard X-ray state} (2001) \emph{MNRAS} {\bf
  327} 1273


\bibitem{tudose06} Tudose V., Fender R. P., Kaiser C., Tzioumis T.,
  van der Klis M., Spencer R., \emph{The large-scale jet-powered radio
    nebula of Circinus X-1} (2006) \emph{MNRAS} {\bf 372} 417
 
\bibitem{watson}
Watson M., Stewart G., King A., Brinkmann W., \emph{Doppler-shifted X-ray line emission from SS~433} (1986) \emph{MNRAS} {\bf 222} 261


\end{thebibliography}
\end{document}